\documentclass[journal]{IEEEtran}
\usepackage{cite}
\usepackage{amsmath,amssymb,amsfonts}
\usepackage{bm}
\usepackage{array}
\usepackage{booktabs}
\usepackage{url}
\usepackage{cases}
\usepackage{graphicx}
\usepackage{stfloats}
\usepackage{subfigure}
\allowdisplaybreaks

\newcommand{\beps}{\boldsymbol{\epsilon}}
\newcommand{\bmu}{\boldsymbol{\mu}}
\newcommand{\bE}{\bm{\mathrm{E}}}
\newcommand{\bH}{\bm{\mathrm{H}}}

\newcommand{\be}{\boldsymbol{e}}
\newcommand{\bh}{\boldsymbol{h}}
\newcommand{\bet}{\boldsymbol{e}_{t}}
\newcommand{\bht}{\boldsymbol{h}_{t}}
\newcommand{\Ett}{\boldsymbol{\epsilon}_{tt}}
\newcommand{\Mtt}{\boldsymbol{\mu}_{tt}}
\newcommand{\etz}{\boldsymbol{\epsilon}_{tz}}
\newcommand{\mtz}{\boldsymbol{\mu}_{tz}}
\newcommand{\Rmat}{\boldsymbol{R}}
\newcommand{\IIt}{\boldsymbol{I}_{2}}
\newcommand{\zh}{\hat{\boldsymbol z}}
\newcommand{\xh}{\hat{\boldsymbol x}}
\newcommand{\yh}{\hat{\boldsymbol y}}
\newcommand{\nh}{\hat{\boldsymbol n}}
\newcommand{\ct}{\nabla_{t}\times}
\newcommand{\dt}{\nabla_{t}\cdot}
\newcommand{\adj}{\operatorname{adj}}

\begin{document}

\title{A Necessary and Sufficient Condition for Complete TE/TM Separability in Fully Anisotropic Waveguides}

\author{Wei Jiang and Jie Liu~\IEEEmembership{Member,~IEEE}

\thanks{Manuscript received XXXX XX, XXXX; revised XXXX XX, XXXX. This work is supported by the National Natural Science Foundation of China under Grant 62561013 and 61901131. \emph{(Corresponding author: Jie Liu)}}
\thanks{W. Jiang is with the School of Physics and Mechatronics Engineering, Guizhou Minzu University, Guiyang 550025, China (e-mail: jwmathphy@163.com)
}
\thanks{J. Liu is with the School of Mathematical Science, Guizhou Normal University, Guiyang 550025, China (e-mail: liujie190484@163.com)
}
}
\mark{Jiang and Liu: Complete TE/TM Separability in Fully Anisotropic Waveguides}

\IEEEtitleabstractindextext{%
\begin{abstract}
A necessary and sufficient condition is established for complete transverse-electric/transverse-magnetic (TE/TM) separability in waveguides homogeneously filled with a fully anisotropic and lossless medium. The permittivity and permeability tensors are Hermitian positive-definite and may contain nonzero transverse--longitudinal coupling terms. Complete TE/TM separation occurs if and only if the transverse permeability block is proportional to the transpose of the transverse permittivity block and the corresponding transverse--longitudinal coupling vectors satisfy the same proportionality relation after complex conjugation. The longitudinal constitutive entries remain independent apart from positive definiteness. The condition is derived directly from the source-free Maxwell equations, and independent analyses of the TE and TM modes lead to the same material relation. Since no boundary condition enters the derivation, the criterion is independent of the boundary type. Full-vector waveguide mode analyses using our in-house programs for perfect-electric-conductor and Bloch-periodic waveguides, with material tensors satisfying and violating the condition, agree with the theoretical results.
\end{abstract}
}
\maketitle
\IEEEdisplaynontitleabstractindextext
\IEEEpeerreviewmaketitle

\section{Introduction}
\IEEEPARstart{T}{ransverse-electric} (TE) and transverse-magnetic (TM) modes are fundamental to the modal description of guided waves. In homogeneous isotropic waveguides, Maxwell's equations naturally separate these two mode classes, whose basic properties are well established in standard treatments of guided waves \cite{Collin1960,Balanis1989}. Electromagnetic propagation in more general anisotropic and inhomogeneous media is discussed in \cite{Chew1990}.

In anisotropic waveguides, coupling among field components generally produces hybrid modes, motivating the development of full-vector numerical formulations. Early finite-element formulations used the electric-field vector to solve electromagnetic and anisotropic waveguide problems \cite{Koshiba1985,Koshiba1986}. Subsequent work introduced tangential-vector elements and efficient full-vector finite-element formulations for electromagnetic field and waveguide computations \cite{Lee1991,Koshiba1992}. Nodal, edge-element, and related formulations were also applied to inhomogeneously loaded and ferrite-filled waveguides \cite{Bardi1993,Zhou1996}. Later studies used edge elements and mixed spectral elements to treat more general anisotropic waveguides \cite{Nuno1997,Liu2015MSEM}. Mixed finite-element and numerical mode-matching formulations have also been used for closed and fully anisotropic waveguide problems \cite{Jiang2018MixedFE,Liu2022NMM}. Recent studies have addressed cutoff and modal computations for homogeneous anisotropic waveguides \cite{Wang2022,Kouroublakis2023}. These methods effectively compute propagation constants and field distributions, but full-vector calculations alone do not provide a constitutive criterion for exact TE and TM separation.

Analytical and numerical studies have also addressed bianisotropic waveguides. Xu and Bosisio derived coupled-mode equations for general bianisotropic waveguides \cite{Xu1995}, while Valor and Zapata developed a finite-element formulation for lossy and inhomogeneous bianisotropic media \cite{Valor1996}. Jakoby and De Zutter used a propagator-matrix formulation to analyze guided waves in inhomogeneous bianisotropic cylindrical structures \cite{Jakoby1996}. In work directly related to TE and TM separation, Graglia \emph{et al.} derived conditions for modal decoupling in homogeneous bianisotropic cylindrical metallic structures and obtained the corresponding longitudinal field equations \cite{Graglia1996}. Uslenghi later reported a concise necessary and sufficient characterization of TE and TM decoupling in a general bianisotropic medium \cite{Uslenghi1997}. For purely anisotropic and lossless waveguides, a necessary and sufficient condition for independent TE and TM modes was established within a restricted transversely anisotropic class \cite{Jiang2017TMTT}. A subsequent study established a sufficient condition for pure TE and TM modes in closed anisotropic waveguides and derived the corresponding scalar and vector eigenvalue formulations \cite{Jiang2018TAP}. More recently, a sufficient condition was obtained and numerically verified for homogeneous fully anisotropic and lossless media with coupling between transverse and longitudinal material components \cite{Jiang2024MOTL}.

This paper establishes a necessary and sufficient constitutive characterization of complete TE and TM separability for the regular modal branches of homogeneous, fully anisotropic, and lossless waveguides with Hermitian positive-definite material tensors. The analysis proceeds directly from the source-free Maxwell equations in the waveguide interior. Neither the theorem nor its proof depends on a specific wall or periodic boundary condition. The material tensors and the chosen propagation direction therefore determine the constitutive criterion, while the boundary condition determines the admissible propagation constants and modal fields.

We derive the necessary condition independently by analyzing the TE and TM cases and show that both lead to the same material requirement. To prove sufficiency, we start from the unreduced coupled longitudinal Maxwell system and establish that the electric and magnetic cross-coupling terms cancel before selecting a pure mode. We then use the separated longitudinal problems to reconstruct the corresponding full Maxwell fields. The analysis thus distinguishes the constitutive requirement for complete TE and TM separability from the spectral admissibility imposed by a particular boundary. Numerical experiments with PEC and Bloch-periodic waveguides examine material tensors that satisfy or violate the condition.

The remainder of this paper is organized as follows. Section~II derives the fully anisotropic waveguide equations from the source-free Maxwell system. Section~III states the main theorem and its corollaries. Section~IV proves necessity by analyzing the TE and TM cases. Section~V proves sufficiency from the coupled longitudinal Maxwell system. Section~VI presents the full-vector modal analysis for PEC and Bloch-periodic waveguides and the numerical verification. Section~VII concludes the paper.

\section{Waveguide Governing Equations}
Let $\Gamma\subset\mathbb R^2$ be the cross section of a $z$-invariant waveguide. Assume that the medium in the waveguide is homogeneous, lossless and fully anisotropic. Hence the permittivity tensor $\beps$ and and permeability tensor $\bmu$ are two constant Hermitian positive-definite matrices. With respect to the transverse plane and the longitudinal direction $\zh$, write
\begin{equation}
\beps=
\begin{bmatrix}
\Ett&\etz\\
\etz^{\dagger}&\epsilon_{zz}
\end{bmatrix},\qquad
\bmu=
\begin{bmatrix}
\Mtt&\mtz\\
\mtz^{\dagger}&\mu_{zz}
\end{bmatrix}.
\label{eq:block-tensors}
\end{equation}
where the symbol $\dagger$ stands for conjugate transpose. Since $\beps$ and $\bmu$ are Hermitian positive-definite, the same
holds for their transverse blocks $\Ett$ and $\Mtt$, which are
therefore nonsingular. Define
\begin{equation*}
\boldsymbol P_{\epsilon}
=
\begin{bmatrix}
\IIt & -\Ett^{-1}\etz\\
\boldsymbol 0^{T} & 1
\end{bmatrix},
\qquad
\boldsymbol P_{\mu}
=
\begin{bmatrix}
\IIt & -\Mtt^{-1}\mtz\\
\boldsymbol 0^{T} & 1
\end{bmatrix},
\end{equation*}
where $\IIt$ is the $2\times2$ identity matrix. Then the congruence transformations give
\begin{equation*}
\boldsymbol P_{\epsilon}^{\dagger}
\beps
\boldsymbol P_{\epsilon}
=
\begin{bmatrix}
\Ett & \boldsymbol 0\\
\boldsymbol 0^{T} &
\epsilon_{zz}
-\etz^{\dagger}\Ett^{-1}\etz
\end{bmatrix},
\end{equation*}
and
\begin{equation*}
\boldsymbol P_{\mu}^{\dagger}
\bmu
\boldsymbol P_{\mu}
=
\begin{bmatrix}
\Mtt & \boldsymbol 0\\
\boldsymbol 0^{T} &
\mu_{zz}
-\mtz^{\dagger}\Mtt^{-1}\mtz
\end{bmatrix}.
\end{equation*}
Since congruence transformations preserve positive definiteness, the corresponding Schur complements are
\begin{equation*}
\epsilon_{zz}
-\etz^{\dagger}\Ett^{-1}\etz>0,
\qquad
\mu_{zz}
-\mtz^{\dagger}\Mtt^{-1}\mtz>0.
\end{equation*}

The time factor in this paper is $\text{e}^{j\omega t}$, where $\omega=2\pi f>0$ is known. A mode propagating in the $+z$ direction is assumed to have the form
\begin{equation}\label{eq:EH}
\bE(x,y,z)=\be(x,y)\text{e}^{-j\beta z},~~
\bH(x,y,z)=\bh(x,y)\text{e}^{-j\beta z},
\end{equation}
where $\beta>0$ is the propagation constant and
\begin{equation}\label{eq:eh}
\be=\bet(x,y)+\zh e_z(x,y),\quad
\bh=\bht(x,y)+\zh h_z(x,y).
\end{equation}
The operator $\nabla$ is decomposed as
\begin{equation}
\nabla=\nabla_t+\zh\frac{\partial}{\partial z},
\quad \nabla_t=\xh\frac{\partial}{\partial x}+\yh\frac{\partial}{\partial y}.
\label{del}
\end{equation}
The source-free Maxwell equations are given by
\begin{subequations}\label{eq:maxwell}
\begin{align}
\nabla\times\bE&=-j\omega\bmu\bH,\\
\nabla\times\bH&= j\omega\beps\bE.
\end{align}
\end{subequations}

Introduce the counterclockwise $90^\circ$ rotation matrix
\begin{equation*}
\Rmat=\begin{bmatrix}0&-1\\1&0\end{bmatrix},
\qquad \Rmat^2=-\IIt,
\end{equation*}
so that $\zh\times\boldsymbol v_t=\Rmat\boldsymbol v_t$ for a transverse vector $\boldsymbol v_t$. Substituting \eqref{eq:block-tensors}--\eqref{del} into \eqref{eq:maxwell} gives
\begin{subequations}
\begin{align}
\nabla_t\times\bet
&=-j\omega\big(\mtz^{\dagger}\bht+\mu_{zz}h_z\big)\zh,
\label{eq:component1}\\
-\Rmat\nabla_t e_z-j\beta\Rmat\bet
&=-j\omega\big(\Mtt\bht+\mtz h_z\big),
\label{eq:component2}\\
\nabla_t\times\bht
&=j\omega\big(\etz^{\dagger}\bet+\epsilon_{zz}e_z\big)\zh,
\label{eq:component3}\\
-\Rmat\nabla_t h_z-j\beta\Rmat\bht
&=j\omega\big(\Ett\bet+\etz e_z\big).
\label{eq:component4}
\end{align}
\end{subequations}

Equations \eqref{eq:component2} and \eqref{eq:component4} form an algebraic system for the transverse fields. Define
\begin{align*}
\boldsymbol f_1&=-\nabla_t e_z-j\omega\Rmat\mtz h_z,\\
\boldsymbol f_2&=j\omega\Rmat\etz e_z-\nabla_t h_z.
\end{align*}
Then
\begin{align*}
j\beta\bet+j\omega\Rmat\Mtt\bht&=\boldsymbol f_1,\\
-j\omega\Rmat\Ett\bet+j\beta\bht&=\boldsymbol f_2.
\end{align*}
Eliminating one transverse field at a time gives
\begin{subequations}\label{eq:N-system}
\begin{align}
\boldsymbol N_1\bet&=\boldsymbol F_1,\\
\boldsymbol N_2\bht&=\boldsymbol F_2,
\end{align}
\end{subequations}
where
\begin{align*}
\boldsymbol N_1&=-\omega^2\Rmat\Mtt\Rmat\Ett-\beta^2\IIt,\\
\boldsymbol N_2&=-\omega^2\Rmat\Ett\Rmat\Mtt-\beta^2\IIt,
\end{align*}
and
\begin{subequations}\label{eq:F-expanded}
\begin{align}
\boldsymbol F_1={}&-j\beta\nabla_t e_z
+\beta\omega\Rmat\mtz h_z \notag\\
&+\omega^2\Rmat\Mtt\Rmat\etz e_z
+j\omega\Rmat\Mtt\nabla_t h_z,
\\
\boldsymbol F_2={}&-j\omega\Rmat\Ett\nabla_t e_z
+\omega^2\Rmat\Ett\Rmat\mtz h_z \notag\\
&-\beta\omega\Rmat\etz e_z
-j\beta\nabla_t h_z.
\end{align}
\end{subequations}
It can be shown that $\boldsymbol N_1$ and $\boldsymbol N_2$ are similar, hence the matrices $\boldsymbol N_1$ and $\boldsymbol N_2$ have the same determinant. When $\det{(\boldsymbol N_1)}=\det(\boldsymbol N_2)\neq0$, it follows from \eqref{eq:N-system} that
\begin{equation}
\bet=\boldsymbol N_1^{-1}\boldsymbol F_1,
\qquad
\bht=\boldsymbol N_2^{-1}\boldsymbol F_2.
\label{eq:tran}
\end{equation}
Substituting \eqref{eq:tran} into \eqref{eq:component1} and \eqref{eq:component3} gives the exact coupled longitudinal equations
\begin{subequations}\label{eq:longitudinal-general}
\begin{align}
\nabla_{t}\times\big(\boldsymbol N_1^{-1}\boldsymbol F_1\big)
&=-j\omega\Big(\mtz^{\dagger}\boldsymbol N_2^{-1}\boldsymbol F_2+\mu_{zz}h_z\Big)\zh,
\label{eq:long1}\\
\nabla_{t}\times\big(\boldsymbol N_2^{-1}\boldsymbol F_2\big)
&=j\omega\Big(\etz^{\dagger}\boldsymbol N_1^{-1}\boldsymbol F_1+\epsilon_{zz}e_z\Big)\zh.
\label{eq:long2}
\end{align}
\end{subequations}
The system \eqref{eq:longitudinal-general} is the starting point for the analysis of necessity and sufficiency. In the following analysis, $\boldsymbol N_1$ and $\boldsymbol N_2$ are assumed to be nonsingular. Therefore, the transverse fields can be eliminated through \eqref{eq:tran}. The singular case is not considered in this work. It may include special modal branches such as TEM modes.

A pure TE mode is defined by
\begin{equation*}
e_z\equiv0,\qquad h_z\not\equiv0\qquad\text{in }\Gamma,
\end{equation*}
whereas a pure TM mode is defined by
\begin{equation*}
h_z\equiv0,\qquad e_z\not\equiv0\qquad\text{in }\Gamma.
\end{equation*}

We next determine which constitutive tensors make the exact longitudinal Maxwell system completely TE/TM separable. Here, complete TE/TM separability means that the longitudinal system \eqref{eq:longitudinal-general} reduces to independent $e_z$ and $h_z$ subproblems, so that a nonzero solution of the corresponding longitudinal boundary-value problem can be reconstructed into a full pure TM or TE mode, respectively. The derivation is based solely on the source-free Maxwell equations and does not involve any boundary condition. The admissible propagation constants are subsequently determined by the specific boundary conditions imposed on the waveguide.

\section{Main Theorem and Consequences}
Let the longitudinal axis of the
homogeneous waveguide be the $z$-axis, and choose the $+z$ direction
as the reference propagation direction. Let $\beps$ and $\bmu$ in \eqref{eq:block-tensors} be constant Hermitian
positive-definite constitutive tensors, where the block decomposition is
taken with respect to the transverse plane orthogonal to $\zh$.

\medskip
\noindent\textbf{Theorem 1 (necessary and sufficient condition):}
For the regular modal branches described above, the souce-free Maxwell system is completely TE/TM separable with respect to the $z$ direction if and only if there exists a real scalar $\rho>0$ such that
\begin{equation}
\Mtt=\rho\Ett^{T},\qquad
\mtz=\rho\overline{\etz},
\label{eq:main-block-condition}
\end{equation}
where $\overline{\etz}$ denotes the complex conjugate of $\etz$. 

The longitudinal entries $\epsilon_{zz}$ and $\mu_{zz}$ are not required to satisfy the same proportionality relation. Equivalently, the block relations can be written as
\begin{equation}
\bmu=\rho\overline{\beps}+\tau\zh\zh^T,
\label{eq:main-compact-condition}
\end{equation}
where $\tau=\mu_{zz}-\rho\epsilon_{zz}$. Positive definiteness of $\bmu$ further requires $\tau>-\rho(\epsilon_{zz}-\etz^{\dagger}\Ett^{-1}\etz)$.
Equation~\eqref{eq:main-compact-condition} shows that the same factor $\rho$ relates all constitutive entries containing at least one transverse index, while the purely longitudinal entries $\epsilon_{zz}$ and $\mu_{zz}$ need not have the same ratio.

\medskip
\noindent\textbf{Corollary 1 (equivalence of the TE and TM material conditions):} The constitutive condition required for TE separation is identical to that required for TM separation. Hence the medium is TE-separable if and only if it is TM-separable at the constitutive level. Whether an individual pure TE or pure TM mode occurs is then determined by the corresponding longitudinal boundary-value problem.

\medskip
\noindent\textbf{Corollary 2 (independence of boundary type):}
The criterion \eqref{eq:main-block-condition} is obtained exclusively from the interior source-free Maxwell equations and does not use a boundary condition. Therefore the same material criterion applies to PEC, PMC, and Bloch-periodic waveguide problems. The boundary condition determines the admissible propagation constants and modal functions, but it does not alter the constitutive relation required for TE/TM separation.

\medskip
\noindent\textbf{Corollary 3 (real reciprocal medium):}
If $\beps$ and $\bmu$ are real symmetric, then
\begin{equation*}
\bmu=\rho\beps+\tau\zh\zh^T.
\end{equation*}
Consequently, after one positive scaling, all entries involving at least one transverse direction coincide and only the $zz$ entry may differ. In particular, the real fully anisotropic family used in \cite{Jiang2024MOTL} is contained in this form.

\medskip
\noindent\textbf{Corollary 4 (transversely anisotropic case):}
If $\etz=\mtz=\boldsymbol 0$, then Theorem~1 reduces to
\begin{equation}
\Mtt=\rho\Ett^T,
\label{eq:transverse-special-case}
\end{equation}
Indeed, if
\begin{equation*}
\Ett=\epsilon
\begin{bmatrix}1&b+cj\\ b-cj&a^2\end{bmatrix},
\end{equation*}
then \eqref{eq:transverse-special-case} gives
\begin{equation*}
\Mtt=\mu
\begin{bmatrix}1&b-cj\\ b+cj&a^2\end{bmatrix},
\qquad \rho=\mu/\epsilon>0,
\end{equation*}
which is precisely the necessary and sufficient condition $\mathfrak C$ of \cite{Jiang2017TMTT}.  Hence the earlier transversely anisotropic theorem is recovered exactly.

\section{Proof of Necessity}
This section proves the necessity of Theorem 1. We first state three elementary lemmas.

\medskip
\noindent\textbf{Lemma 1:}
Let $\boldsymbol X\in\mathbb C^{2\times2}$ be a nonzero constant matrix. Then
\begin{equation*}
\nabla_{t}\times(\boldsymbol X\nabla_t\psi)=0
\quad\text{for every }\psi\in C^2(\Gamma)
\end{equation*}
if and only if $\boldsymbol X=\lambda\IIt$ for some nonzero scalar $\lambda$.

\emph{Proof:} See Lemma 1 in \cite{Jiang2017TMTT}  \hfill$\square$

\medskip
\noindent\textbf{Lemma 2:}
For every $\boldsymbol X\in\mathbb C^{2\times2}$,
\begin{equation*}
-\Rmat\boldsymbol X^T\Rmat=\adj(\boldsymbol X).
\end{equation*}
If $\boldsymbol X$ is invertible,
\begin{equation*}
-\Rmat\boldsymbol X^{-1}\Rmat
=\frac{\boldsymbol X^T}{\det\boldsymbol X}.
\end{equation*}

\emph{Proof:} Both identities follow by direct multiplication. \hfill$\square$

\medskip
\noindent\textbf{Lemma 3:}
The matrix $-\omega^2\Rmat\Mtt\Rmat\Ett$ is a scalar matrix if and only if
$-\omega^2\Rmat\Ett\Rmat\Mtt$ is a scalar matrix. Moreover, if
either of them is a scalar matrix, then there exists a real scalar
$k^2>0$ such that
\begin{equation}
-\omega^2\Rmat\Mtt\Rmat\Ett
=
-\omega^2\Rmat\Ett\Rmat\Mtt
=
k^2\IIt .
\label{eq:positive-scalar-wavenumber}
\end{equation}

\emph{Proof:}
According to Theorem~1 in \cite{Jiang2017TMTT},
$-\omega^2\Rmat\Mtt\Rmat\Ett$ is a scalar matrix if and only if
$-\omega^2\Rmat\Ett\Rmat\Mtt$ is a scalar matrix. Furthermore,
whenever this condition holds,
\begin{equation}
\Rmat\Mtt\Rmat\Ett
=
\Rmat\Ett\Rmat\Mtt .
\label{eq:two-products-equal}
\end{equation}
It therefore remains only to prove that the common scalar in
\eqref{eq:positive-scalar-wavenumber} is real and strictly positive.

Define
\begin{equation}\label{eq:A-definition}
  \boldsymbol A:=-\Rmat\Mtt\Rmat.
\end{equation}
Since $\Rmat^\dagger=-\Rmat$, we have $\boldsymbol A
=\Rmat^\dagger\Mtt\Rmat$. For every nonzero vector $\boldsymbol x\in\mathbb C^2$,
\begin{align*}
\boldsymbol x^\dagger\boldsymbol A\boldsymbol x
&=
\boldsymbol x^\dagger
\Rmat^\dagger\Mtt\Rmat\boldsymbol x
=
(\Rmat\boldsymbol x)^\dagger
\Mtt
(\Rmat\boldsymbol x)
>0 .
\end{align*}
Here $\Rmat\boldsymbol x\neq\boldsymbol 0$ because $\Rmat$ is
nonsingular. Hence, $\boldsymbol A$ is Hermitian positive-definite
and is therefore nonsingular.

Suppose now that
$-\Rmat\Mtt\Rmat\Ett$ is a scalar matrix, i.e., there exists a scalar $\eta\in\mathbb C$ such that
\begin{equation}
-\Rmat\Mtt\Rmat\Ett
=
\eta\IIt .
\label{eq:eta-scalar}
\end{equation}
Using \eqref{eq:A-definition}, the relation \eqref{eq:eta-scalar} can be written as
\begin{equation}
\boldsymbol A\Ett=\eta\IIt .
\label{eq:AEt-eta}
\end{equation}
Since $\boldsymbol A$ is nonsingular, left-multiplying
\eqref{eq:AEt-eta} by $\boldsymbol A^{-1}$ gives
\begin{equation}
\Ett=\eta\boldsymbol A^{-1}.
\label{eq:Et-Ainverse}
\end{equation}
Right-multiplying \eqref{eq:Et-Ainverse} by $\boldsymbol A$ yields
\begin{equation}
\Ett\boldsymbol A=\eta\IIt .
\label{eq:EtA-eta}
\end{equation}
Combining \eqref{eq:AEt-eta} and \eqref{eq:EtA-eta}, we obtain
\begin{equation*}
\boldsymbol A\Ett
=
\Ett\boldsymbol A
=
\eta\IIt .
\end{equation*}
Thus, $\boldsymbol A$ and $\Ett$ commute.

Both $\boldsymbol A$ and $\Ett$ are Hermitian positive-definite, and since they commute, their product $\boldsymbol A\Ett$ is also
Hermitian positive-definite. Therefore, for every nonzero
$\boldsymbol x\in\mathbb C^2$, $\boldsymbol x^\dagger
\boldsymbol A\Ett
\boldsymbol x>0$. On the other hand, by \eqref{eq:AEt-eta},
$
\boldsymbol x^\dagger
\boldsymbol A\Ett
\boldsymbol x
=
\eta\,\boldsymbol x^\dagger\boldsymbol x$. Consequently, $\eta =\frac{
\boldsymbol x^\dagger
\boldsymbol A\Ett
\boldsymbol x
}{
\boldsymbol x^\dagger\boldsymbol x
}
>0$. Hence, $\eta$ is necessarily real and strictly positive.

Finally, define $k^2=\omega^2\eta$. Since $\omega>0$ and $\eta>0$, it follows that $k^2>0$. Equation~\eqref{eq:eta-scalar} then gives
\begin{equation*}
-\omega^2\Rmat\Mtt\Rmat\Ett
=\omega^2\eta\IIt=
k^2\IIt .
\end{equation*}
Together with \eqref{eq:two-products-equal}, this further yields
\begin{equation*}
-\omega^2\Rmat\Ett\Rmat\Mtt
=\omega^2\eta\IIt=
k^2\IIt .
\end{equation*}
Therefore, \eqref{eq:positive-scalar-wavenumber} holds.
\hfill$\square$

\subsection{Necessity obtained from the TE mode}
For a TE mode, set $e_z\equiv0$ in \eqref{eq:longitudinal-general}. Then we have
\begin{subequations}
\begin{align}
\nabla_{t}\times\big(\boldsymbol N_1^{-1}\boldsymbol F_{1}^{\rm TE}\big)
&=-j\omega\Big(\mtz^{\dagger}\boldsymbol N_2^{-1}\boldsymbol F_{2}^{\rm TE}+\mu_{zz}h_z\Big)\zh,
\label{eq:temodes1}\\
\nabla_{t}\times\big(\boldsymbol N_2^{-1}\boldsymbol F_{2}^{\rm TE}\big)
&=j\omega\Big(\etz^{\dagger}\boldsymbol N_1^{-1}\boldsymbol F_{1}^{\rm TE}\Big)\zh.
\label{eq:temodes2}
\end{align}
\end{subequations}
where $\boldsymbol F_1^{\rm TE}$ and $\boldsymbol F_2^{\rm TE}$ are defined by

\begin{align*}
\boldsymbol F_1^{\rm TE}
&=\beta\omega\Rmat\mtz h_z
+j\omega\Rmat\Mtt\nabla_t h_z,\\
\boldsymbol F_2^{\rm TE}
&=\omega^2\Rmat\Ett\Rmat\mtz h_z
-j\beta\nabla_t h_z.
\end{align*}
\indent The equation \eqref{eq:temodes1} is the nontrivial longitudinal TE equation. The equation \eqref{eq:temodes2} must be compatible with the TE reduction and must not impose an additional independent longitudinal equation. Therefore equation \eqref{eq:temodes2} must hold identically for the TE longitudinal field.

The only second-order term in \eqref{eq:temodes2} is $-j\beta\ct\big(\boldsymbol N_2^{-1}\nabla_t h_z\big)$ and
all remaining terms contain at most first derivatives of $h_z$. Therefore
\begin{equation*}
  -j\beta\ct\big(\boldsymbol N_2^{-1}\nabla_t h_z\big)=0,\quad{\rm in}~\Gamma
\end{equation*}
Since $\beta>0$, Lemma~1 gives $\boldsymbol N_2=n\IIt$. Since $\boldsymbol N_2=-\omega^2\Rmat\Ett\Rmat\Mtt-\beta^2\IIt$, $-\omega^2\Rmat\Ett\Rmat\Mtt$ is a scalar matrix. From Lemma 3, there exists a positive scalar $k^2>0$ such that
\begin{equation*}
-\omega^2\Rmat\Ett\Rmat\Mtt=k^2\IIt.
\end{equation*}
Using Lemma~2 and the positive definiteness of $\Ett$ and $\Mtt$ yields
\begin{equation}
\Mtt=\rho\Ett^T,
\qquad
\rho=\frac{k^2}{\omega^2\det\Ett}>0.
\label{eq:M-from-TE}
\end{equation}
From Lemma 3, we can define
\begin{equation}
\boldsymbol N_1=\boldsymbol N_2=k_t^2\IIt, ~
k_t^2=k^2-\beta^2, ~
k^2=\omega^2\rho\det(\Ett).
\label{eq:N-same}
\end{equation}
In this case, $k$ is called the medium wavenumber in the anisotropic lossless medium and
$k_{t}$ is called the corresponding cut-off wavenumber. When $-\omega^2\Rmat\Ett\Rmat\Mtt$ or $-\omega^2\Rmat\Mtt\Rmat\Ett$ is not a scalar matrix,
it is impossible to define the medium wavenumber of such a fully anisotropic medium.

The TE compatibility equation \eqref{eq:temodes2} reduces to
\begin{equation*}
(\Ett\Rmat\mtz)^T\nabla_t h_z
=-\etz^{\dagger}\Rmat\Mtt\nabla_t h_z
+j\beta\etz^{\dagger}\Rmat\mtz\,h_z.
\end{equation*}
The coefficients of $\nabla_t h_z$ and $h_z$ must vanish separately.
Thus we have
\begin{subequations}
\begin{align}
(\Ett\Rmat\mtz)^T&=-\etz^{\dagger}\Rmat\Mtt,
\label{eq:TE-first}\\
\etz^{\dagger}\Rmat\mtz&=0.
\label{eq:TE-zero}
\end{align}
\end{subequations}
Substituting \eqref{eq:M-from-TE} into \eqref{eq:TE-first} gives
\begin{equation*}
-\mtz^T\Rmat\Ett^T=-\rho\etz^{\dagger}\Rmat\Ett^T.
\end{equation*}
Since $\Rmat\Ett^T$ is nonsingular, $\mtz^T=\rho\etz^{\dagger}$ holds,
and hence
\begin{equation*}
\mtz=\rho\overline{\etz}.
\end{equation*}
Equation \eqref{eq:TE-zero} is then satisfied automatically because $\etz^{\dagger}\Rmat\overline{\etz}=\overline{\etz^T\Rmat\etz}=0$. Thus the TE definition alone forces the two relations in \eqref{eq:main-block-condition}.

\subsection{Necessity obtained from a TM mode}
For a TM mode, set $h_z\equiv0$ in \eqref{eq:longitudinal-general}. Then we have
\begin{subequations}
\begin{align}
\nabla_{t}\times\big(\boldsymbol N_1^{-1}\boldsymbol F_{1}^{\rm TM}\big)
&=-j\omega\Big(\mtz^{\dagger}\boldsymbol N_2^{-1}\boldsymbol F_{2}^{\rm TM}\Big)\zh,
\label{eq:tmmodes1}\\
\nabla_{t}\times\big(\boldsymbol N_2^{-1}\boldsymbol F_{2}^{\rm TM}\big)
&=j\omega\Big(\etz^{\dagger}\boldsymbol N_1^{-1}\boldsymbol F_{1}^{\rm TM}+\epsilon_{zz}e_z\Big)\zh.
\label{eq:tmmodes2}
\end{align}
\end{subequations}
where $\boldsymbol F_1^{\rm TM}$ and $\boldsymbol F_2^{\rm TM}$ are defined by
\begin{align*}
\boldsymbol F_1^{\rm TM}
&=-j\beta\nabla_{t}e_z
+\omega^2\Rmat\Mtt\Rmat\etz e_z,\\
\boldsymbol F_2^{\rm TM}
&=-j\omega\Rmat\Ett\nabla_{t}e_z
-\beta\omega\Rmat\etz e_z.
\end{align*}

The equation \eqref{eq:tmmodes2} is the nontrivial longitudinal TM equation. The equation \eqref{eq:tmmodes1} must be compatible with the TM reduction and must not impose an additional independent longitudinal equation. Therefore equation \eqref{eq:tmmodes1} must hold identically for the TM longitudinal field.

The only second-order term in \eqref{eq:tmmodes1} is $-j\beta\,\ct\big(\boldsymbol N_1^{-1}\nabla_t e_z\big)$ and
all remaining terms contain at most first derivatives of $e_z$. Therefore
\begin{equation*}
-j\beta\,\ct\big(\boldsymbol N_1^{-1}\nabla_t e_z\big)=0, \quad{\rm in}~\Gamma
\end{equation*}
so Lemma~1 gives $\boldsymbol N_1=n\IIt$. Exactly as above,
\begin{equation*}
\Mtt=\rho\Ett^T,
\qquad
\boldsymbol N_1=\boldsymbol N_2=k_t^2\IIt.
\end{equation*}

The TM compatibility equation \eqref{eq:tmmodes1} reduces to
\begin{equation*}
(\Mtt\Rmat\etz)^T\nabla_t e_z
=-\mtz^{\dagger}\Rmat\Ett\nabla_t e_z
+j\beta\mtz^{\dagger}\Rmat\etz\,e_z.
\end{equation*}
Hence
\begin{align*}
(\Mtt\Rmat\etz)^T&=-\mtz^{\dagger}\Rmat\Ett,\\
\mtz^{\dagger}\Rmat\etz&=0.
\end{align*}

Using $\Mtt=\rho\Ett^T$ in the first identity again gives
\begin{equation*}
\mtz=\rho\overline{\etz}.
\end{equation*}
Therefore, the independent analyses of the TE and TM cases lead to exactly the same material condition. This proves the necessity and Corollary~1.

\section{Proof of Sufficiency}
This section proves the sufficiency of Theorem 1. Assume that
\begin{equation}
\Mtt=\rho\Ett^T,
\qquad
\mtz=\rho\overline{\etz},
\qquad \rho>0.
\label{eq:suff-assumption}
\end{equation}
We prove that, whenever the corresponding longitudinal boundary-value problem admits a nonzero solution, that solution generates a full Maxwell mode for which the other longitudinal field component vanishes identically. Thus the vanishing longitudinal component is a consequence of the Maxwell equations and the material condition.

By Lemma~2,
\begin{equation*}
-\Rmat\Mtt\Rmat\Ett
=-\Rmat\Ett\Rmat\Mtt
=\rho\det(\Ett)\IIt.
\end{equation*}
Consequently \eqref{eq:N-same} holds. Split the forcing terms in \eqref{eq:F-expanded} according to their longitudinal component:
\begin{align*}
\boldsymbol F_1&=\boldsymbol F_{1e}+\boldsymbol F_{1h},\\
\boldsymbol F_2&=\boldsymbol F_{2e}+\boldsymbol F_{2h},
\end{align*}
with
\begin{subequations}
\begin{align}
\boldsymbol F_{1e}
&=-j\beta\nabla_t e_z-k^2\Ett^{-1}\etz\,e_z,
\label{eq:F1e}\\
\boldsymbol F_{1h}
&=\beta\omega\Rmat\mtz h_z
+j\omega\Rmat\Mtt\nabla_t h_z,
\\
\boldsymbol F_{2e}
&=-j\omega\Rmat\Ett\nabla_t e_z
-\beta\omega\Rmat\etz e_z,
\\
\boldsymbol F_{2h}
&=\omega^2\Rmat\Ett\Rmat\mtz h_z
-j\beta\nabla_t h_z.
\end{align}
\end{subequations}
Here \eqref{eq:F1e} follows from
$\omega^2\Rmat\Mtt\Rmat=-k^2\Ett^{-1}$.

Multiplying \eqref{eq:long1}--\eqref{eq:long2} by $k_t^2$ gives
\begin{subequations}\label{eq:longitudinal-scaled}
\begin{align}
\zh\cdot\ct\boldsymbol F_1+j\omega\mtz^{\dagger}\boldsymbol F_2
+j\omega\mu_{zz}k_t^2h_z&=0,
\label{eq:longscaled1}\\
\zh\cdot\ct\boldsymbol F_2-j\omega\etz^{\dagger}\boldsymbol F_1
-j\omega\epsilon_{zz}k_t^2e_z&=0.
\end{align}
\end{subequations}
We first prove that the two longitudinal components in \eqref{eq:longitudinal-scaled} are uncoupled. For the $e_z$-dependent part of \eqref{eq:longscaled1}, use
$\mtz^{\dagger}=\rho\etz^T$, $\etz^T\Rmat\etz=0$, and the $2\times2$ identity
\begin{equation*}
(\det\Ett)\Ett^{-T}\Rmat=\Rmat\Ett.
\end{equation*}
A direct calculation gives
\begin{equation}
\zh\cdot\ct\boldsymbol F_{1e}
+j\omega\mtz^{\dagger}\boldsymbol F_{2e}=0.
\label{eq:cross-e-cancel}
\end{equation}
Similarly,
\begin{equation}
\zh\cdot\ct\boldsymbol F_{2h}
-j\omega\etz^{\dagger}\boldsymbol F_{1h}=0.
\label{eq:cross-h-cancel}
\end{equation}
Therefore the general coupled longitudinal system \eqref{eq:longitudinal-scaled} reduces to
\begin{subequations}
\begin{align}
\zh\cdot\ct\boldsymbol F_{1h}
+j\omega\mtz^{\dagger}\boldsymbol F_{2h}
+j\omega k_t^2\mu_{zz}h_z&=0,
\label{eq:h-only-compact}\\
\zh\cdot\ct\boldsymbol F_{2e}
-j\omega\etz^{\dagger}\boldsymbol F_{1e}
-j\omega k_t^2\epsilon_{zz}e_z&=0.
\label{eq:e-only-compact}
\end{align}
\end{subequations}
Expanding these two equations gives
\begin{align}
\dt(\Mtt\nabla_t h_z)
&-j\beta(\mtz^T+\mtz^{\dagger})\nabla_t h_z \notag\\
&+\Big[\mu_{zz}k_t^2
-k^2\mtz^{\dagger}\Mtt^{-1}\mtz\Big]h_z=0,
\label{eq:TE-scalar}
\end{align}
and
\begin{align}
\dt(\Ett\nabla_t e_z)
&-j\beta(\etz^T+\etz^{\dagger})\nabla_t e_z \notag\\
&+\Big[\epsilon_{zz}k_t^2
-k^2\etz^{\dagger}\Ett^{-1}\etz\Big]e_z=0.
\label{eq:TM-scalar}
\end{align}
We now prove constructively that a nonzero solution of \eqref{eq:TE-scalar} generates a pure TE mode, whereas a nonzero solution of \eqref{eq:TM-scalar} generates a pure TM mode.

\subsection{Construction of a pure TE mode}
Let $h_z\not\equiv0$ satisfy \eqref{eq:TE-scalar}, together with the boundary condition induced by the chosen waveguide boundary. Define the transverse fields by
\begin{subequations}\label{eq:TE-reconstruction}
\begin{align}
\bet^{\rm TE}
&:=\frac{1}{k_t^2}\boldsymbol F_{1h}
=\frac{\beta\omega}{k_t^2}\Rmat\mtz h_z
+\frac{j\omega}{k_t^2}\Rmat\Mtt\nabla_t h_z,
\\
\bht^{\rm TE}
&:=\frac{1}{k_t^2}\boldsymbol F_{2h}
=\frac{\omega^2}{k_t^2}\Rmat\Ett\Rmat\mtz h_z
-\frac{j\beta}{k_t^2}\nabla_t h_z.
\end{align}
\end{subequations}
At this stage, no value has been prescribed for the longitudinal electric component. It is recovered from the longitudinal part of Maxwell equation \eqref{eq:component3}:
\begin{equation}
\widehat e_z
:=\frac{1}{\epsilon_{zz}}
\left[
\frac{1}{j\omega}\zh\cdot\ct\bht^{\rm TE}
-\etz^{\dagger}\bet^{\rm TE}
\right].
\label{eq:TE-recovered-ez}
\end{equation}
Substituting \eqref{eq:TE-reconstruction} into \eqref{eq:TE-recovered-ez} yields
\begin{align}
\widehat e_z
=\frac{1}{\epsilon_{zz}k_t^2}
\left[
\frac{1}{j\omega}\zh\cdot\ct\boldsymbol F_{2h}
-\etz^{\dagger}\boldsymbol F_{1h}
\right]=0,
\label{eq:TE-ez-zero}
\end{align}
where the last equality follows exactly from \eqref{eq:cross-h-cancel}. Hence the longitudinal electric component generated by the full Maxwell system vanishes identically.

It remains to verify the other Maxwell equations. Equation \eqref{eq:h-only-compact}, divided by $k_t^2$, gives
\begin{equation*}
\zh\cdot\ct\bet^{\rm TE}
=-j\omega\left(
\mtz^{\dagger}\bht^{\rm TE}+\mu_{zz}h_z
\right),
\end{equation*}
which is precisely \eqref{eq:component1}. The two transverse Maxwell equations also follow directly from the reconstruction. Indeed,
\begin{align*}
&~~~~k_t^2\left(
 j\beta\bet^{\rm TE}+j\omega\Rmat\Mtt\bht^{\rm TE}
\right) \notag\\
&=j\beta\boldsymbol F_{1h}
+j\omega\Rmat\Mtt\boldsymbol F_{2h} =-j\omega k_t^2\Rmat\mtz h_z,
\end{align*}
where $-\omega^2\Rmat\Mtt\Rmat\Ett=k^2\IIt$ has been used. Therefore
\begin{equation}
 j\beta\bet^{\rm TE}+j\omega\Rmat\Mtt\bht^{\rm TE}
=-j\omega\Rmat\mtz h_z.
\label{eq:TE-transverse1}
\end{equation}
Similarly,
\begin{align*}
&~~~~k_t^2\left(
-j\omega\Rmat\Ett\bet^{\rm TE}+j\beta\bht^{\rm TE}
\right) \notag\\
&=-j\omega\Rmat\Ett\boldsymbol F_{1h}
+j\beta\boldsymbol F_{2h} =-k_t^2\nabla_t h_z,
\end{align*}
so that
\begin{equation}
-j\omega\Rmat\Ett\bet^{\rm TE}+j\beta\bht^{\rm TE}
=-\nabla_t h_z.
\label{eq:TE-transverse2}
\end{equation}
Equations \eqref{eq:TE-transverse1} and \eqref{eq:TE-transverse2} are exactly \eqref{eq:component2} and \eqref{eq:component4} with the value \eqref{eq:TE-ez-zero} obtained from Maxwell's equations. Consequently,
\begin{equation*}
\bE^{\rm TE}=\bet^{\rm TE}e^{-j\beta z},
\qquad
\bH^{\rm TE}=(\bht^{\rm TE}+\zh h_z)e^{-j\beta z}
\end{equation*}
is a full Maxwell mode and satisfies
\begin{equation*}
\widehat e_z\equiv0,
\qquad h_z\not\equiv0.
\end{equation*}
Thus it is a pure TE mode.

\subsection{Construction of a pure TM mode}
Let $e_z\not\equiv0$ satisfy \eqref{eq:TM-scalar}, together with the boundary condition induced by the chosen waveguide boundary. Define
\begin{subequations}\label{eq:TM-reconstruction}
\begin{align}
\bet^{\rm TM}
&:=\frac{1}{k_t^2}\boldsymbol F_{1e}
=-\frac{j\beta}{k_t^2}\nabla_t e_z
-\frac{k^2}{k_t^2}\Ett^{-1}\etz e_z,
\\
\bht^{\rm TM}
&:=\frac{1}{k_t^2}\boldsymbol F_{2e}
=-\frac{j\omega}{k_t^2}\Rmat\Ett\nabla_t e_z
-\frac{\beta\omega}{k_t^2}\Rmat\etz e_z.
\end{align}
\end{subequations}
The longitudinal magnetic component is again not prescribed. It is recovered from \eqref{eq:component1}:
\begin{equation*}
\widehat h_z
:=\frac{1}{\mu_{zz}}
\left[
-\frac{1}{j\omega}\zh\cdot\ct\bet^{\rm TM}
-\mtz^{\dagger}\bht^{\rm TM}
\right].
\end{equation*}
Using \eqref{eq:TM-reconstruction} and \eqref{eq:cross-e-cancel},
\begin{align}
\widehat h_z
&=\frac{1}{\mu_{zz}k_t^2}
\left[
-\frac{1}{j\omega}\zh\cdot\ct\boldsymbol F_{1e}
-\mtz^{\dagger}\boldsymbol F_{2e}
\right] \notag\\
&=0.
\label{eq:TM-hz-zero}
\end{align}
Thus the longitudinal magnetic component is forced to vanish by the Maxwell equations and the material condition.

Equation \eqref{eq:e-only-compact}, divided by $k_t^2$, gives
\begin{equation*}
\zh\cdot\ct\bht^{\rm TM}
=j\omega\left(
\etz^{\dagger}\bet^{\rm TM}+\epsilon_{zz}e_z
\right),
\end{equation*}
which is \eqref{eq:component3}. For the transverse equations,
\begin{align*}
&~~~~k_t^2\left(
 j\beta\bet^{\rm TM}+j\omega\Rmat\Mtt\bht^{\rm TM}
\right) \notag\\
&=j\beta\boldsymbol F_{1e}
+j\omega\Rmat\Mtt\boldsymbol F_{2e} =-k_t^2\nabla_t e_z,
\end{align*}
and therefore
\begin{equation}
 j\beta\bet^{\rm TM}+j\omega\Rmat\Mtt\bht^{\rm TM}
=-\nabla_t e_z.
\label{eq:TM-transverse1}
\end{equation}
Likewise,
\begin{align*}
&~~~~k_t^2\left(
-j\omega\Rmat\Ett\bet^{\rm TM}+j\beta\bht^{\rm TM}
\right) \notag\\
&=-j\omega\Rmat\Ett\boldsymbol F_{1e}
+j\beta\boldsymbol F_{2e} =j\omega k_t^2\Rmat\etz e_z,
\end{align*}
which gives
\begin{equation}
-j\omega\Rmat\Ett\bet^{\rm TM}+j\beta\bht^{\rm TM}
=j\omega\Rmat\etz e_z.
\label{eq:TM-transverse2}
\end{equation}
Equations \eqref{eq:TM-transverse1} and \eqref{eq:TM-transverse2} are exactly \eqref{eq:component2} and \eqref{eq:component4} with the value \eqref{eq:TM-hz-zero} obtained from Maxwell's equations. Hence
\begin{equation*}
\bE^{\rm TM}=(\bet^{\rm TM}+\zh e_z)e^{-j\beta z},
\qquad
\bH^{\rm TM}=\bht^{\rm TM}e^{-j\beta z}
\end{equation*}
is a full Maxwell mode satisfying
\begin{equation*}
\widehat h_z\equiv0,
\qquad e_z\not\equiv0.
\end{equation*}
Thus it is a pure TM mode.

These constructions prove sufficiency. More precisely, after the material condition \eqref{eq:suff-assumption} has been imposed, every nonzero solution of the TE longitudinal boundary-value problem \eqref{eq:TE-scalar} generates a full Maxwell solution for which the longitudinal electric component is identically zero, and every nonzero solution of the TM longitudinal boundary-value problem \eqref{eq:TM-scalar} generates a full Maxwell solution for which the longitudinal magnetic component is identically zero. Therefore the material condition in Theorem~1 is sufficient for pure TE and pure TM modes.

\medskip
\noindent\textbf{Remark 1:}
The constitutive condition is independent of the boundary type. A specific PEC, PMC, or Bloch-periodic boundary condition only determines whether the corresponding scalar longitudinal problem admits a nonzero eigenfunction and which values of $\beta$ are admissible. Once such an eigenfunction exists, the above reconstruction proves that the associated full Maxwell field is a pure TE or pure TM mode.

\section{Numerical Experiments}

The numerical experiments in this section serve only to verify the theoretical condition established in Theorem~1. The computations are performed with our in-house codes on triangular meshes. The transverse electric field $\bet$ is discretized by the LT/QN tangential-vector element, whose tangential component is linear and normal component is quadratic, while the longitudinal component $e_z$ is approximated by quadratic Lagrange elements. This mixed full-vector discretization follows the same general approach as mixed finite-element waveguide formulations, and its algorithmic details are not repeated here. The solver directly solves the full-vector eigenproblem for the electric field $\bE$. The magnetic field $\bH$ is then reconstructed from Maxwell's equations, and the modal type is identified from $e_z$ and $h_z$. Thus the appearance of a TE, TM, or hybrid mode is a numerical consequence of the full Maxwell system and is not imposed a priori.

For the full-vector waveguide mode analysis, define $\nabla_{\beta}=\nabla_t-j\beta\zh$. Using the relative material tensors $\beps_r$ and $\bmu_r$, the electric-field eigenproblem is
\begin{equation}
\nabla_{\beta}\times
\left(\bmu_r^{-1}\nabla_{\beta}\times\be\right)
=k_0^2\beps_r\be
\qquad\text{in }\Gamma,
\label{eq:numerical-full-model}
\end{equation}
where the wavenumber in vacuum $k_0=\omega\sqrt{\epsilon_0\mu_0}$ is known, and the modal electric field $\be=\bet+\zh e_z$ and  the propagation constant $\beta$ are unknown. Once $\be$ is computed, the modal magnetic field is reconstructed as $\bh=(j/\omega\mu_0)\bmu_r^{-1}(\nabla_{\beta}\times\be)$. A computed mode is then classified according to whether the longitudinal components $e_z$ and $h_z$ vanish. 

The cross section of the waveguide is chosen as the rectangle
\begin{equation*}
\Gamma=(0,L_x)\times(0,L_y),
\end{equation*}
with
\begin{equation*}
L_x=15~\text{cm},
\qquad
L_y=10~\text{cm}.
\end{equation*}
In all numerical experiments, the operating frequency is
$0.5~\text{GHz}$.

\subsection{Waveguide configurations for mode analysis}

To distinguish the present fully anisotropic theory from the
real-symmetric sufficient family considered in \cite{Jiang2024MOTL},
the numerical examples use genuinely complex Hermitian material
tensors.

The first material satisfies the condition in Theorem~1:
\begin{eqnarray}
\beps_r^{(S)}
&=&
\begin{bmatrix}
2&1+j&1+j\\
1-j&4&1-j\\
1-j&1+j&5
\end{bmatrix},
\label{eq:material-S1}\\
\bmu_r^{(S)}
&=&
\begin{bmatrix}
4&2-2j&2-2j\\
2+2j&8&2+2j\\
2+2j&2-2j&12
\end{bmatrix}.
\label{eq:material-S2}
\end{eqnarray}
Both tensors are Hermitian and fully anisotropic, with nonzero complex
$xy$, $xz$, and $yz$ coupling terms. Their transverse blocks and
transverse--longitudinal coupling vectors satisfy
\begin{equation*}
\Mtt^{(S)}
=
2\big(\Ett^{(S)}\big)^T,
\qquad
\mtz^{(S)}
=
2\overline{\etz^{(S)}}.
\end{equation*}
Therefore, for the corresponding absolute constitutive tensors, the
condition in Theorem~1 holds with $\rho=2\mu_0/\epsilon_0$. Positive definiteness can be checked directly from the Schur
complements:
\begin{equation*}
\epsilon_{zz}^{(S)}
-
\big(\etz^{(S)}\big)^{\dagger}
\big(\Ett^{(S)}\big)^{-1}
\etz^{(S)}
=
\frac{11}{3}>0,
\end{equation*}
and
\begin{equation*}
\mu_{zz}^{(S)}
-
\big(\mtz^{(S)}\big)^{\dagger}
\big(\Mtt^{(S)}\big)^{-1}
\mtz^{(S)}
=
\frac{28}{3}>0.
\end{equation*}
Also,
\begin{equation*}
\mu_{zz}^{(S)}
=
12
\neq
2\epsilon_{zz}^{(S)}
=
10.
\end{equation*}
Thus this example is not restricted to the special full
proportionality between the two constitutive tensors. It
illustrates the freedom of the $zz$ entry permitted by Theorem~1.

The second material is chosen to violate only the
transverse--longitudinal relation while preserving the same complex
transverse blocks. Its relative permeability tensor is
\begin{equation}
\bmu_r^{(V)}
=
\begin{bmatrix}
4&2-2j&2.4-2j\\
2+2j&8&2+2j\\
2.4+2j&2-2j&12
\end{bmatrix},
\label{eq:material-V}
\end{equation}
whereas its relative permittivity tensor is the same as
\eqref{eq:material-S1}. Hence
\begin{equation*}
\Mtt^{(V)}
=
2\big(\Ett^{(V)}\big)^T,
\end{equation*}
but
\begin{equation*}
\mtz^{(V)}
=
\begin{bmatrix}
2.4-2j\\
2+2j
\end{bmatrix}
\neq
2\overline{\etz^{(V)}}
=
\begin{bmatrix}
2-2j\\
2+2j
\end{bmatrix}.
\end{equation*}
The violating permeability tensor remains Hermitian positive definite;
indeed,
\begin{equation*}
\mu_{zz}^{(V)}
-
\big(\mtz^{(V)}\big)^{\dagger}
\big(\Mtt^{(V)}\big)^{-1}
\mtz^{(V)}
=
\frac{676}{75}>0.
\end{equation*}
Thus the second material violates only the second relation in
Theorem~1, while all the other assumptions of the theorem are
retained. The two complex material sets therefore provide a
positive/negative test of the necessary and sufficient
condition.

For the PEC waveguide, \eqref{eq:numerical-full-model} is supplemented
by the PEC boundary condition
\begin{equation*}
\nh\times\bE=0
\qquad\text{on }\partial\Gamma.
\end{equation*}
For the modal representation, this condition is equivalent on the
cross-sectional boundary to
\begin{equation*}
\nh\times\bet=0,
\qquad
e_z=0
\qquad\text{on }\partial\Gamma.
\end{equation*}

For the Bloch-periodic waveguide,
\eqref{eq:numerical-full-model} is supplemented by
\begin{align*}
\bet(L_x,y)
&=
\text{e}^{-jk_xL_x}\bet(0,y),
&&0<y<L_y,\\
\bet(x,L_y)
&=
\text{e}^{-jk_yL_y}\bet(x,0),
&&0<x<L_x,\\
e_z(L_x,y)
&=
\text{e}^{-jk_xL_x}e_z(0,y),
&&0<y<L_y,\\
e_z(x,L_y)
&=
\text{e}^{-jk_yL_y}e_z(x,0),
&&0<x<L_x,
\end{align*}
where
\begin{equation*}
k_x=k_0\sin\theta\cos\phi,
\qquad
k_y=k_0\sin\theta\sin\phi,
\end{equation*}
with
\begin{equation*}
\theta=\frac{\pi}{6},
\qquad
\phi=\frac{\pi}{4}.
\end{equation*}

\subsection{Four numerical experiments}

Four numerical experiments are considered, corresponding to the two
material tensors under PEC and Bloch-periodic boundary conditions.
For each configuration, the full-vector Maxwell eigenproblem is solved
without imposing either $e_z=0$ or $h_z=0$. Representative computed
modes are shown below. Here, we list the total degrees of freedom $N_{\text{dof}}$ resulting from numerical discretization.

\begin{figure}[!t]
  \centering
  \includegraphics[width=0.95\columnwidth,draft=false]{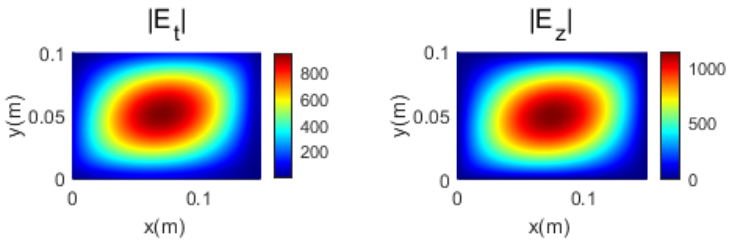}
  \includegraphics[width=0.95\columnwidth,draft=false]{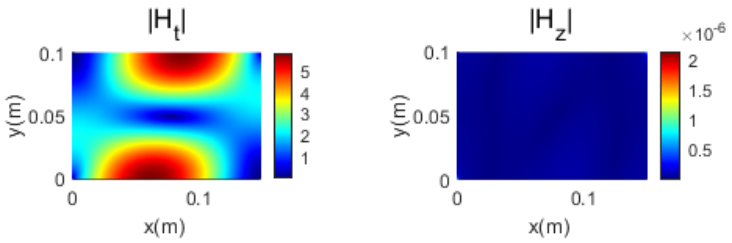}
  \caption{Electromagnetic field distribution of a TM mode in the PEC
  waveguide filled with the first material,
  $\beta\approx4.3204~\text{rad/m}$ and $N_{\rm dof}=83196$.}
  \label{fig:pec-S-TM}
\end{figure}

\begin{figure}[!t]
  \centering
  \includegraphics[width=0.95\columnwidth,draft=false]{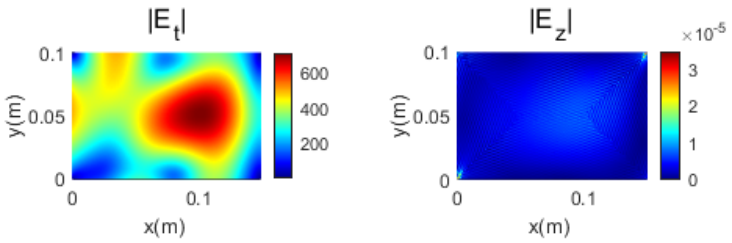}
  \includegraphics[width=0.95\columnwidth,draft=false]{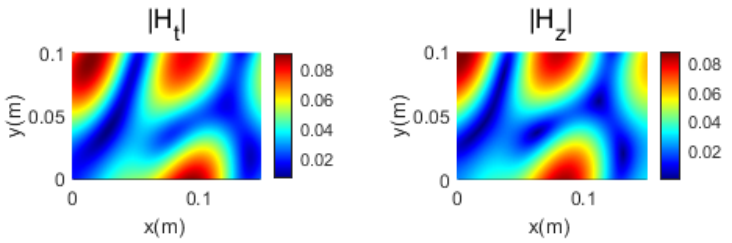}
  \caption{Electromagnetic field distribution of a TE mode in the PEC
  waveguide filled with the first material,
  $\beta\approx13.4609~\text{rad/m}$ and $N_{\rm dof}=83196$.}
  \label{fig:pec-S-TE}
\end{figure}

For the PEC waveguide filled with the first material
\eqref{eq:material-S1}--\eqref{eq:material-S2}, the computed spectrum
contains pure TM and pure TE modes. Figure~\ref{fig:pec-S-TM} shows a
representative TM mode with
$\beta\approx4.3204~\text{rad/m}$, whereas
Fig.~\ref{fig:pec-S-TE} shows a representative TE mode with
$\beta\approx13.4609~\text{rad/m}$.

\begin{figure}[!t]
  \centering
  \includegraphics[width=0.95\columnwidth,draft=false]{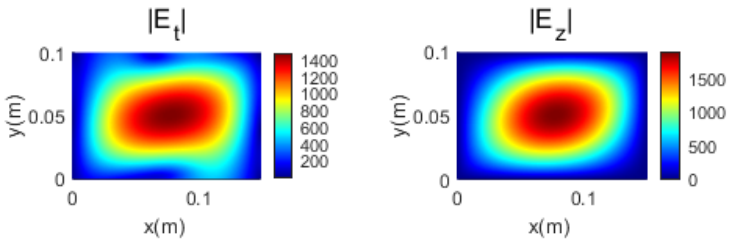}
  \includegraphics[width=0.95\columnwidth,draft=false]{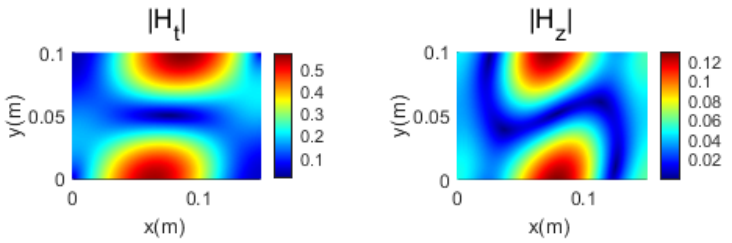}
  \caption{Electromagnetic field distribution of the first hybrid mode
  in the PEC waveguide filled with the second material,
  $\beta\approx12.8140~\text{rad/m}$ and $N_{\rm dof}=83196$.}
  \label{fig:pec-V-H1}
\end{figure}

\begin{figure}[!t]
  \centering
  \includegraphics[width=0.95\columnwidth,draft=false]{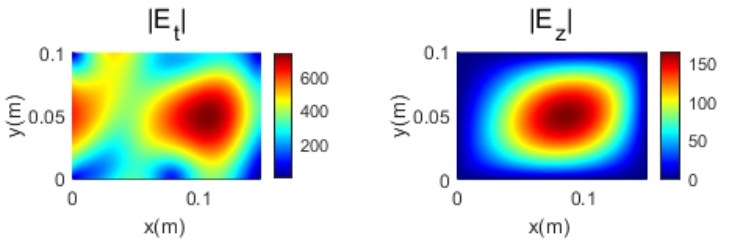}
  \includegraphics[width=0.95\columnwidth,draft=false]{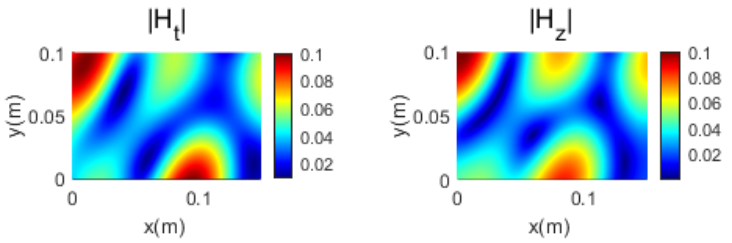}
  \caption{Electromagnetic field distribution of the second hybrid mode
  in the PEC waveguide filled with the second material,
  $\beta\approx14.3523~\text{rad/m}$ and $N_{\rm dof}=83196$.}
  \label{fig:pec-V-H2}
\end{figure}

In contrast, when the PEC waveguide is filled with the second material
\eqref{eq:material-S1} and \eqref{eq:material-V}, which violates the constitutive condition in
Theorem~1, the computed modes are hybrid. Two representative hybrid
modes are shown in Figs.~\ref{fig:pec-V-H1} and
\ref{fig:pec-V-H2}. In both cases, neither longitudinal field component
vanishes, so the modes cannot be classified as pure TE or pure TM.

\begin{figure}[!t]
  \centering
  \includegraphics[width=0.95\columnwidth,draft=false]{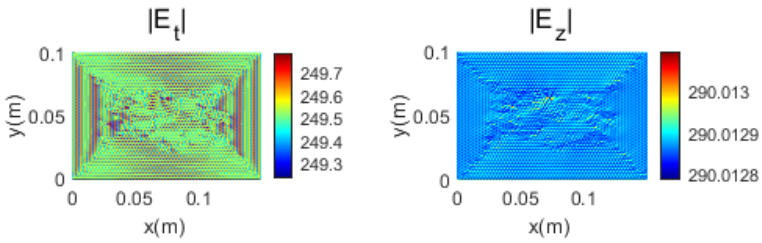}
  \includegraphics[width=0.95\columnwidth,draft=false]{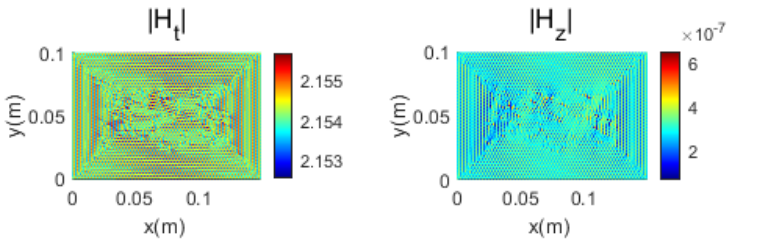}
  \caption{Electromagnetic field distribution of a TM mode in the
  Bloch-periodic waveguide filled with the first material,
  $\beta\approx2.5317~\text{rad/m}$ and $N_{\rm dof}=27600$.}
  \label{fig:bloch-S-TM}
\end{figure}

\begin{figure}[!t]
  \centering
  \includegraphics[width=0.95\columnwidth,draft=false]{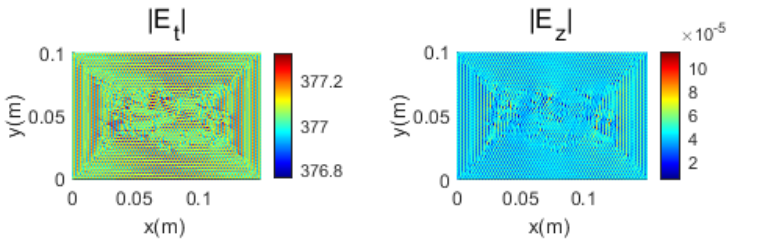}
  \includegraphics[width=0.95\columnwidth,draft=false]{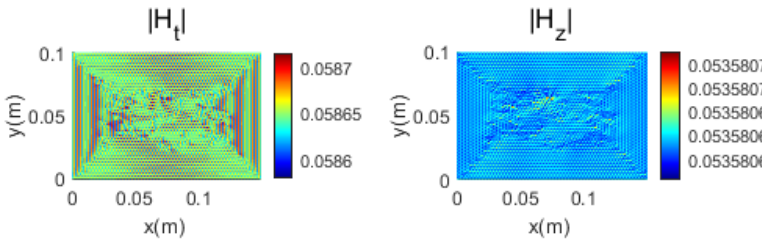}
  \caption{Electromagnetic field distribution of a TE mode in the
  Bloch-periodic waveguide filled with the first material,
  $\beta\approx10.0615~\text{rad/m}$ and $N_{\rm dof}=27600$.}
  \label{fig:bloch-S-TE}
\end{figure}

The same distinction is observed under Bloch-periodic boundary
conditions. When the first material is used, the full-vector
calculation again produces pure TM and TE modes. A representative TM
mode with $\beta\approx2.5317~\text{rad/m}$ is shown in
Fig.~\ref{fig:bloch-S-TM}, and a representative TE mode with
$\beta\approx10.0615~\text{rad/m}$ is shown in
Fig.~\ref{fig:bloch-S-TE}.

\begin{figure}[!t]
  \centering
  \includegraphics[width=0.95\columnwidth,draft=false]{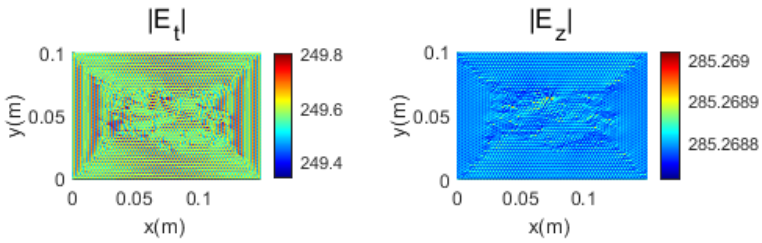}
  \includegraphics[width=0.95\columnwidth,draft=false]{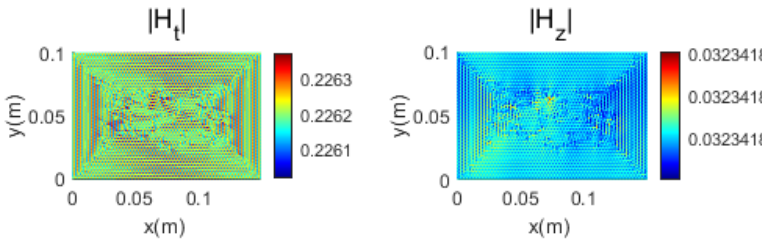}
  \caption{Electromagnetic field distribution of the first hybrid mode
  in the Bloch-periodic waveguide filled with the second material,
  $\beta\approx2.1307~\text{rad/m}$ and $N_{\rm dof}=27600$.}
  \label{fig:bloch-V-H1}
\end{figure}

\begin{figure}[!t]
  \centering
  \includegraphics[width=0.95\columnwidth,draft=false]{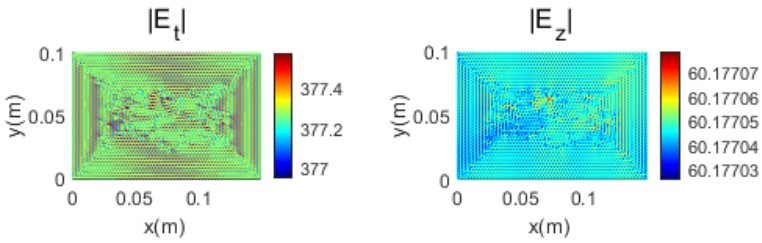}
  \includegraphics[width=0.95\columnwidth,draft=false]{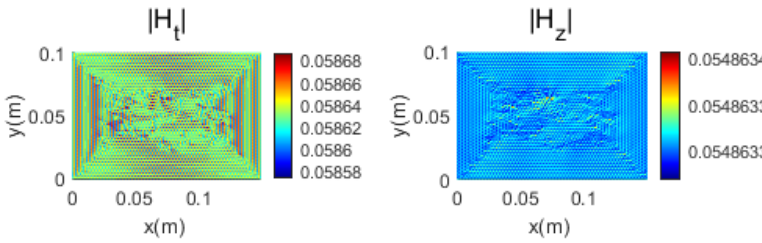}
  \caption{Electromagnetic field distribution of the second hybrid mode
  in the Bloch-periodic waveguide filled with the second material,
  $\beta\approx8.6530~\text{rad/m}$ and $N_{\rm dof}=27600$.}
  \label{fig:bloch-V-H2}
\end{figure}

Finally, when the second material is used in the Bloch-periodic
waveguide, the representative computed modes are again hybrid.
Figures~\ref{fig:bloch-V-H1} and \ref{fig:bloch-V-H2} show two such
modes with propagation constants
$\beta\approx2.1307~\text{rad/m}$ and
$\beta\approx8.6530~\text{rad/m}$, respectively.

The numerical results show the same qualitative behavior under the two boundary conditions. The first material satisfies \eqref{eq:main-block-condition}. The full-vector calculations then produce pure TE and pure TM modes for both PEC and Bloch-periodic waveguides. The second material violates only the transverse--longitudinal constitutive relation. The representative computed modes are hybrid under both boundary conditions.

These results are consistent with Theorem~1 and Corollary~2. The first material shows that the constitutive condition permits complete TE/TM separation. The second material shows that the separation is lost when the condition is violated. The same behavior is observed for both PEC and Bloch-periodic waveguides. This result supports the boundary independence of the constitutive criterion. The material tensors determine the TE/TM separability. The boundary condition determines the admissible propagation constants and modal field distributions.

\section{Conclusion}
A necessary and sufficient material condition has been derived for complete TE/TM separability in a homogeneous, fully anisotropic, and lossless waveguide. The derivation starts from the source-free Maxwell equations. Forward propagation along the $+z$ direction is assumed, with the time convention $\text{e}^{j\omega t}$. The transverse field components are then eliminated. This gives the exact coupled longitudinal system.

The necessity proof uses only the direct physical definitions: $e_z\equiv0$ for TE modes and $h_z\equiv0$ for TM modes. Each reduction independently yields the same relations
\begin{equation*}
\Mtt=\rho\Ett^T,
\qquad
\mtz=\rho\overline{\etz},
\qquad \rho>0.
\end{equation*}
Thus TE and TM modes have the same material condition. For the sufficiency proof, the material relations are substituted into the full coupled longitudinal equations. The electric-to-magnetic and magnetic-to-electric cross terms then cancel identically. The system separates into an $e_z$ equation and an $h_z$ equation. The corresponding pure TM and pure TE modes are then reconstructed from these decoupled equations.

No boundary condition is used in the theorem or its proof. Therefore, the material criterion is independent of the boundary type. It applies to PEC, PMC, and Bloch-periodic waveguides. For numerical verification, full-vector waveguide mode analyses are performed for PEC and Bloch-periodic waveguides. An in-house MATLAB finite-element code is used. One material satisfies the proposed condition, while the other violates it. The modal type is identified according to whether the longitudinal components $e_z$ and $h_z$ vanish.

The obtained condition can be used in practical waveguide analysis and design. It provides a direct test of whether a fully anisotropic medium permits complete TE/TM separation. The condition can be checked from the constitutive tensors before a full-vector mode analysis is performed. It can also guide the selection of material parameters when pure TE or TM modes are required. Since the criterion is independent of the boundary type, it can be applied to different waveguide boundary configurations.

\end{document}